\documentclass[reprint,prb,superscriptaddress]{revtex4-2}

\usepackage{color}
\usepackage{amsmath,amsthm}
\usepackage{graphicx}
\usepackage{multirow}

\newcommand{\av}[1]{\left\langle #1 \right \rangle}

\newcommand{\up}{\uparrow}
\newcommand{\dn}{\downarrow}

\begin{document}

\title{Static treatment of dynamic interactions in the single-orbital Anderson impurity model}

\author{Anton Pauli}
\affiliation{NanoLund and Division of Mathematical Physics,
    Department of Physics,
    Lund University, Lund,
    Sweden}

\author{Akshat Mishra}
\affiliation{NanoLund and Division of Mathematical Physics,
    Department of Physics,
    Lund University, Lund,
    Sweden}

\author{Malte Rösner}
\affiliation{Institute for Molecules and Materials, Radboud University, Nijmegen, The Netherlands}

\author{Erik G. C. P. van Loon}
\affiliation{NanoLund and Division of Mathematical Physics,
    Department of Physics,
    Lund University, Lund,
    Sweden}
    
\begin{abstract}
Correlated electron physics is intrinsically a multiscale problem, since high-energy electronic states screen the interactions between the correlated electrons close to the Fermi level, thereby reducing the magnitude of the interaction strength and dramatically shortening its range. Thus, the handling of screening is an essential ingredient in the first-principles modelling of correlated electron systems. 
Screening is an intrinsically dynamic process and the corresponding downfolding methods such as the constrained Random Phase Approximation indeed produce a dynamic interaction. However, many low-energy methods require an instantaneous interaction as input, which makes it necessary to map the fully dynamic interaction to an effective instantaneous interaction strength. It is a priori not clear if and when such an effective model can capture the physics of the one with dynamic interaction and how to best perform the mapping. Here, we provide a systematic benchmark relevant to correlated materials, in the form of the Anderson impurity model. Overall, we find that a static approximation can be valid and that the moment-based approach recently proposed by Scott and Booth can be a good tool to find the value of the static interaction. We also identify physical regimes, especially under doping, where an instantaneous interaction cannot capture all of the relevant physics.    
\end{abstract}

\maketitle

\section{Introduction}

Quantum materials are characterized by potentially strong and complicated correlations between their electrons, which results in a high sensitivity to the details of the electronic interaction~\cite{basov2017towards}. An example of this sensitivity is the appearance of competing phases with tiny energy differences~\cite{xu2024coexistence}. This sensitivity is the key to tunability and the promise of properties on demand in quantum materials, and therefore to their technological applications. However, the sensitivity also poses a challenge to first-principle modelling, since high accuracy across energy scales is necessary. 

To address the multiscale nature of the problem, first-principles modelling of correlated electrons typically starts by integrating out weakly correlated degrees of freedom using relatively simple methods, to end up with a low-energy model with fewer degrees of freedom that can be solved using more advanced quantum many-body methods. The goal of this downfolding procedure~\cite{Aryasetiawan2022} is therefore to determine the correct low-energy model, both in terms of the structure of the model and in the values of the parameters.  

Concretely, the state-of-the-art for strongly correlated systems is to perform downfolding starting from the band structure according to density functional theory or $GW$ theory~\cite{Boehnke16,tomczak2017merging,choi2016first}, mapping it to a low-energy model solved using, e.g., dynamical mean-field theory (DMFT)~\cite{Georges96,Kotliar06}, with the interactions in the low-energy model determined using the constrained Random Phase Approximation (cRPA)~\cite{Aryasetiawan04}. 

The unscreened Coulomb interaction is instantaneous (in the non-relativistic limit) and long-ranged, but the account of screening in the downfolding process leads to a short-ranged, non-instantaneous interaction~\cite{Aryasetiawan04,Werner16}. The introduction of retardation effects is common whenever degrees of freedom are integrated out, and is also present in methods such as constrained Functional Renormalization Group (cFRG)~\cite{Honerkamp12}.

\begin{figure}
\includegraphics{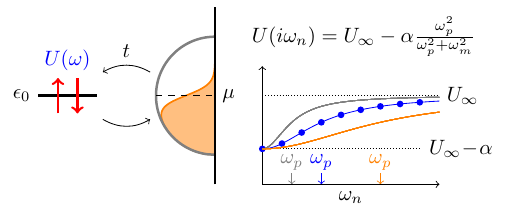}
\caption{Left: Single-impurity Anderson model with dynamic interaction $U(\omega)$. Right: Dynamic interaction in the Matsubara representation. The three colored lines correspond to interactions with the same $U_\infty$ and $\alpha$ but different $\omega_p$ (indicated by small vertical arrows). Due to this parametrization, all three curves have the same values at $n=0$ and $n\rightarrow \infty$. }
\label{fig:SIAM}
\end{figure}
Practically, however, most many-body approaches prefer to work with a low-energy model with instantaneous interactions. This raises the question of how to pick an instantaneous interaction that appropriately captures the physics of the underlying dynamic interaction. A common but ad hoc choice is to take the low-frequency limit of the dynamic interaction, under the assumption that the screening processes are much faster (higher energy) than the low-energy physics of the correlated electrons. It has been observed that this simple approach often leads to too low values~\cite{Shinaoka15,Honerkamp18,vanLoon21,carta2025} of the interaction in DFT+cRPA+DMFT, when comparing the final results to experiment, but it is hard to assess if this mismatch arises due to the neglect of dynamic interactions or due to any of the many other approximations in the computational chain. Other possible culprits are the choice of Wannier functions, the neglect of vertex corrections in cRPA, the treatment of double-counting corrections and the fact that DMFT is not the exact solution of the downfolded model~\cite{Georges96,Rohringer18}.

Recently, Scott and Booth have proposed  an alternative recipe to determine the effective instantaneous interaction, which they call mRPA (moment RPA)~\cite{BoothMRPA}. They benchmarked this approach in the context of molecules using quantum chemistry methods, while the applicability of mRPA to the case of correlated solids within the context of DMFT and its extensions~\cite{Rohringer18} has not yet been established. In particular, the relevant physics of correlated metals is not present in the original benchmark using molecules.

Here we study the role of dynamic versus instantaneous interactions in exactly solvable quantum impurity models, giving a rigorous benchmark to answer two questions: (1) Is it at all possible to use an instantaneous interaction parameter to appropriately describe the full, dynamic interaction? (2) Does the mRPA provide a good approximate value of the instantaneous interaction? To this end, we use a single-orbital Anderson impurity model coupled to a continuous bath, with a dynamic interaction $U(\omega)$ on the impurity, as sketched in Fig.~\ref{fig:SIAM}. This model is exactly solvable using continuous-time Monte Carlo~\cite{Werner07,Gull:2011lr}, is sign-problem free even away from half-filling, and captures the essential physics of local electronic correlations in solids in a controlled set-up.

\begin{table}
\caption{Overview of electronic bandwidth $W$, interaction and screening parameters in selected materials. The bandwidth $W$ is that of the low-energy space and has been estimated from the density of states or band structure given in the references. For multi-orbital models, the bare ($U_\infty$) and screened $U_{\omega=0}$ values given here are approximate averages over the density-density elements. These values are given as a way to judge the orders of magnitude, the precise values are sensitive to the DFT+cRPA implementation and the choice of Wannier orbitals~\cite{Reddy25}. Parameters for additional materials can be found in Ref.~\cite{Casula12}.}
\begin{tabular}{|p{2cm}|c|c|c|c|c|} 
 \hline
 Material & $W$ (eV) & $U_\infty$ (eV) & $U_{i\omega_0}$ (eV) & $\omega_p$ (eV) & Ref. \\ [0.5ex] 
 \hline\hline
 CrI$_3$  & 1.0 & 13 &  3.5 & 22.0 & \cite{Soriano2021} \\ 
 \hline
 SrVO$_3$ & 1.4 & 16.5 & 3.3 & 18.0 & \cite{Casula12} \\ 
% \hline
% Sr$_2$VO$_4$ & 1.4 & 15.7 & 3.1 & 18.1 & \cite{Casula12} \\ 
% \hline
% LaVO$_3$ & 1.1 & 13.3 & 1.9 & 10.3 & \cite{Casula12} \\ 
% \hline
% VO$_2$ & 1.3 & 15.2 & 2.7 & 15.6 & \cite{Casula12} \\ 
% \hline
% TaS$_2$ & 1.6 & 8.4 & 1.5 & 14.7 & \cite{Casula12} \\ 
% \hline
% SrMnO$_3$ & 1.0 & 21.6 & 3.1 & 13.3 & \cite{Casula12} \\ 
% \hline
% BaFe$_2$As$_2$ & 1.2 & 19.7 & 2.8 & 15.7 & \cite{Casula12} \\ 
% \hline
% LaOFeAs & 1.2 & 19.1 & 2.7 & 16.5 & \cite{Casula12} \\ 
% \hline
% FeSe & 1.3 & 20.7 & 4.2 & 17.4 & \cite{Casula12} \\ 
% \hline
% CuO & 1.3 & 26.1 & 6.8 & 21.1 & \cite{Casula12} \\ 
 \hline
 Ni & 4.0 & 24.0 & 3.5 & 27.0 & \cite{Aryasetiawan04, BlugelPRL2012, MiyakePRB2009}\\ 
 \hline
 Cr & 4.0 & 19.0 & 4.0 & 20.0 & \cite{BlugelPRL2012}\\ 
 \hline
 Graphite & 16.8 & 17.6 & 8.0 & 20.0 & \cite{WehlingPRL2011}\\
 \hline
 Graphene & 16.8 & 17.0 & 9.3 & 20.0 & \cite{WehlingPRL2011}\\
 \hline
Nb$_3$Cl$_8$ \newline monolayer & 0.2 & 7.0 & 2.0 & 25.0 & \cite{grytsiuk2024nb3cl8}\\ 
 \hline
UO$_2$ ($f$ only) & 2 & 16 & 3.5 & 8.2, 16.3  & \cite{Amadon14} \\
\hline
% Benzene \EvL{Units?} % Units of Hartree?
%& 10 & \EvL{4.2} & \EvL{3.5} & \EvL{1.7} & \cite{ScottPRL2024,WehlingPRL2011} \\
% \hline
Vanadocene & 2.2  & 13 & 5 & 16.5 & \cite{chang2024downfolding} \\
\hline
\end{tabular}
\label{tab:values}
\end{table}

\begin{figure}
\includegraphics{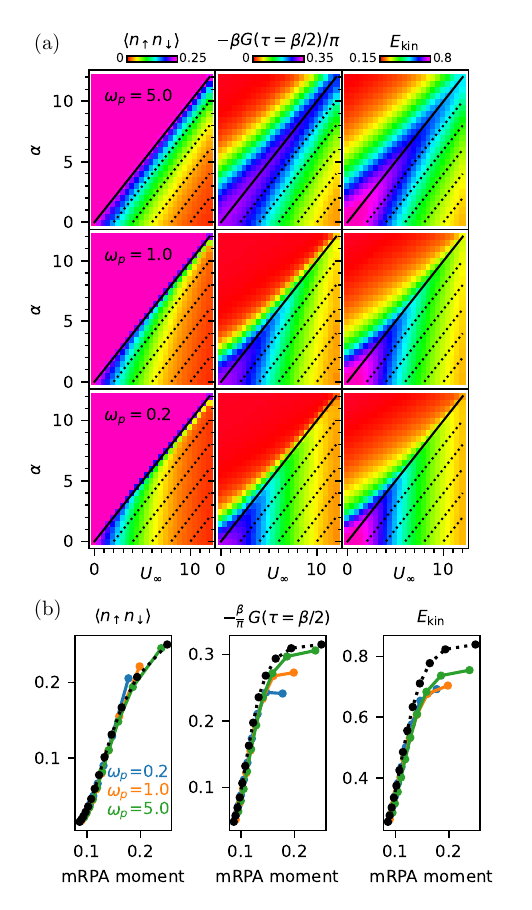}
\caption{Observables in the half-filled system at $\beta=10$. (a) Color maps in the $U_\infty-\alpha$ plane, the black diagonal line shows $U_\infty-\alpha=U(i\omega_0)=0$ and the dotted lines are parallel to this line. The color scale for the double occupancy is restricted to show more detail in the region of interest. (b) Observables as a function of mRPA moment, Eq.~\eqref{eq:mrpa:moment}, at $\alpha=3$. Note that a larger $U_\infty$ corresponds to a smaller mRPA moment, so the correlation strength increases going to the left in this plot. $\alpha=0$ reference results are shown in black. The mRPA works whenever the colored and black lines are on top of each other.}
\label{fig:halffilled:colormap:beta10}
\end{figure} 

\section{Model} 
The impurity model is studied in equilibrium at finite temperature $T=1/\beta$, so that we work with the Matsubara representation of the dynamic interaction, $U(i\omega_n)$ with $\omega_n=2\pi n/\beta$. A single-pole model for the dynamic interaction is used, 
\begin{align}
 U(i\omega_n) = U_\infty - \alpha \frac{\omega_p^2}{\omega_p^2+\omega_n^2}. \label{eq:Uw}
\end{align}
This model corresponds to a real-energy interaction with a single pole at $\omega_p$, the plasma frequency. $U_\infty=U(i\omega_{n}\rightarrow \infty)$ is the unscreened value of the interaction and $\alpha$ is the strength of the dynamic screening, i.e., $\alpha = U_\infty-U(i\omega_0)$.  Such a single-pole model of screening~\cite{lundqvist1967a,lundqvist1967b,Hybertsen85,Aryasetiawan98} is frequently used in the context of computational methods such as $GW$, while the more general case involves multiple pole frequencies and screening strengths, $U(i\omega_n)=U_\infty-\sum_j \alpha_j \frac{\omega_{p,j}^2}{\omega_{p,j}^2+\omega_n^2}$. Multiple poles can arise from inter-band and intra-band processes involving different bands~\cite{Amadon14}.

Apart from the dynamic interaction, the impurity model is characterized by the coupling $t=1$ to the bath and the bath density of states, which is chosen to be semi-circular with bandwidth 4, which sets the overall energy scale of the problem.
We perform calculations at fixed density $n=1$ (half-filling) and $n=0.8$ (hole-doped), modifying the chemical potential and the impurity potential $\epsilon_0$ to reach the desired density both on the impurity and in the bath.
The model has $SU(2)$ spin-symmetry, so the spin-averaged Green's functions and densities are used to reduce the Monte Carlo noise. We use units of $\hbar=1$ and $k_B=1$. 

Altogether, the action of the impurity model is
\begin{align}
 S &= -\int_0^\beta d\tau \int_0^\beta d\tau' \sum_{\sigma=\uparrow,\downarrow} c^*_\sigma \mathcal{G}_{0}^{-1}(\tau-\tau') c_\sigma \notag \\
&+ \frac{1}{2} \int_0^\beta d\tau \int_0^\beta d\tau'  \,\, n(\tau) U(\tau-\tau') n(\tau'),  
\end{align}
where $U(\tau-\tau')$ is the Fourier transform of Eq.~\eqref{eq:Uw} and $\mathcal{G}_0(\tau-\tau')$ is the Fourier transform of $\mathcal{G}^{-1}_{0}(i\nu_n) = i\nu_n+\mu-\epsilon_0-t^2 G^\text{bath}(i\nu_n)$, where $G^\text{bath}$ is the Matsubara Green's function corresponding to a semicircular density of states of width $W=4$.

Relevant parameter values for selected materials are shown in Table~\ref{tab:values}. $U(i\omega_0)$ is typically much smaller than $U_\infty$, i.e., $\alpha$ is generally large. For the strongly correlated materials, $W$ is small so $\omega_p/W$ is often at least 5 if not larger. Exceptions are graphite and graphene, which have a large bandwidth so that $W\approx \omega_p$, but which also feature only weak correlations. 

The impurity model is solved using the CTSEG solver~\cite{Kavokine2025} available with the TRIQS toolbox~\cite{triqs}, TPRF~\cite{Strand:tprf} is used for the data processing. CTSEG is a continuous-time hybridization expansion solver~\cite{Werner:2006rt,Gull:2011lr} that can deal with dynamic interactions~\cite{Werner07,Hafermann13,Hafermann14b}. The solver is sign-problem-free for the single-orbital model. In the parameter regime studied here, the cost for a single data point is of the order of a core-minute and mostly depends on temperature. The code is available at Ref.~\cite{zenodo}. 

To investigate different temporal aspects of the electronic correlations, our benchmark is based on three observables: the double occupancy $\av{n_\up n_\dn}$, the kinetic energy $E_\text{kin}$, and the middle point of the imaginary time Green's function $-\frac{\beta}{\pi}G(\tau=\beta/2)$. The double occupancy is measured directly in the solver and is an indicator of the energy associated with the instantaneous part of the interaction. The kinetic energy is related to the measured average perturbation order $\av{k}$ in the impurity solver~\cite{Gull:2011lr} as $E_\text{kin}=\av{k}/\beta$ and shows the amount of delocalization of the electrons. Note that we have defined $E_\text{kin}$ so that it is positive. Finally, $-\frac{\beta}{\pi}G(\tau=\beta/2)$ gives an estimate of the density of states around the Fermi level~\cite{Ayral13,vanLoon14}.

\section{mRPA}

Scott and Booth~\cite{BoothMRPA} propose to choose the effective static interaction $U^\ast$ so that the equal-time RPA susceptibility is equal to that in the model with dynamic interactions
\begin{align}
 \chi^{\text{RPA},U^\ast}(\tau=0) &\overset{!}{=} \chi^{\text{RPA},U(i\omega_n)}(\tau=0). \label{eq:mRPA}
\end{align}
A similar mapping to static interactions based on equal-time correlation functions has been considered for the dual boson approach~\cite{Peters19}.
The equal-time susceptibility corresponds to moments of the dynamic susceptibility (hence the name Moment RPA) or equivalently to elements of the two-particle density matrix. For the paramagnetic single-orbital model considered here, the only non-trivial element of this density matrix is the double occupancy $\av{n_\up n_\dn}$, since the other elements are fixed by the electron density. Similarly, $U^\ast$ is a scalar, while it is generally a tensor in multi-orbital models. Since the RPA susceptibility can be calculated analytically even in the presence of dynamic interactions, it is numerically straightforward to find the solution $U^\ast$ for Eq.~\eqref{eq:mRPA}. More details and explicit formulas are provided in Appendix~\ref{sec:app:mrpa}.

\begin{figure*}[!ht]
\includegraphics{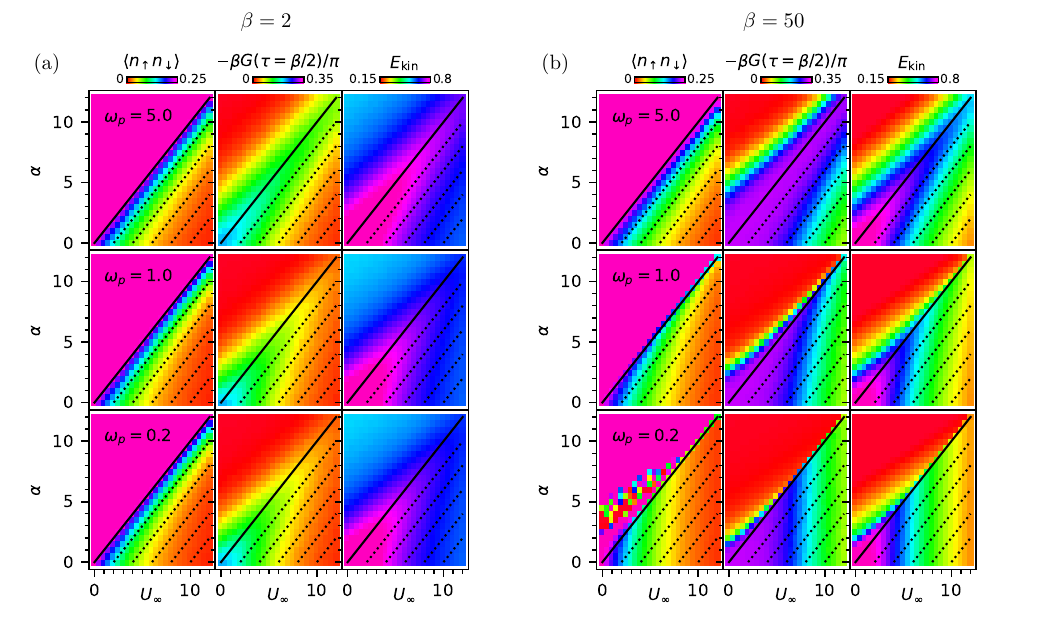}
\caption{Observables in the half-filled system at $\beta=2$ (a) and $\beta=50$ (b). The color scales are the same as in Fig.~\ref{fig:halffilled:colormap:beta10}.}
\label{fig:halffilled:colormap:beta2}
\end{figure*}

\section{Results at half-filling} 
Figure~\ref{fig:halffilled:colormap:beta10}(a) shows several observables in the $U_\infty$-$\alpha$ plane, for a fixed temperature $\beta=10$. The color map is chosen so that the shape of isolines is easy to spot. 

For $\alpha=0$, i.e., with a purely instantaneous interaction, increasing $U_\infty$ lowers the double occupancy, kinetic energy and $-\frac{\beta}{\pi}G(\tau\!\!=\!\!\beta/2)$, which corresponds to the evolution from a weakly correlated to a strongly correlated regime. The evolution is smooth since we are looking at a single impurity model which does not have any genuine phase transitions. 

The color maps show that the increase in $\alpha$ clearly counteracts the effect of $U_\infty$ at half-filling. For large $\alpha$, i.e., $\alpha \gtrsim U_\infty$ (black line), the interaction at low frequency becomes attractive and the impurity model enters the bipolaron regime~\cite{shubin1935elektronentheorie,Vonsovsky79}, where the impurity is either empty or doubly-occupied. Thus, the double occupancy exceeds the non-interacting value $\frac{1}{4}$ (the color map is cut off at this value). This regime of the impurity model occurs in Extended DMFT~\cite{Ayral13} for the Extended Hubbard model in the charge-density wave phase~\cite{Hirsch84}.

Although all three values of $\omega_p$ are globally similar, there are also clear differences. For $\omega_p=5$, the double occupancy is approximately constant at constant $U_\infty-\alpha$, i.e., along the dotted lines, showing that the simple formula $U^\ast=U(i\omega_{0})=U_\infty-\alpha$ works well for this observable. This makes sense when looking at $U(i\omega_n)$ in Fig.~\ref{fig:SIAM}, since large $\omega_p$ means that $U(i\omega_n)$ remains close to the zero-frequency value over a large range of Matsubara frequencies. Deviations between the isolines of the colormap and the dotted lines are clearly visible at the lower values of $\omega_p$. For example, for the double occupancy, the yellow isoline crosses several dotted lines at both $\omega_p=1$ and $\omega_p=0.2$. For the kinetic energy and Green's function, although competition between $U_\infty$ and $\alpha$ is visible, the picture is less ``diagonal'' than that of the double occupancy. 

Figure~\ref{fig:halffilled:colormap:beta2} shows the equivalent of Fig.~\ref{fig:halffilled:colormap:beta10} but at higher and lower temperature, respectively. At lower temperature, the dependence of the slope of the isolines (constant color) on $\omega_p$ is even clearer, going from diagonal to almost vertical. Furthermore, features in the colormap become sharper at lower temperature, as expected from statistical physics. The double occupancy in the bipolaron regime ($\alpha>U_\infty$) is showing clear Monte Carlo noise, but the more relevant $\alpha<U_\infty$ regime remains numerically stable. Generally, the CTSEG Monte Carlo simulations are more challenging at larger $\beta$ (lower temperature). In summary, the results show that $\alpha$ counteracts $U_\infty$ in a similar way for all observables, so that a $U^\ast$ model could succeed, but that the ad hoc choice $U^\ast=U(i\omega_0)$ is clearly insufficient for intermediate and small $\omega_p$.

This brings us to the second question: How well does the mRPA value of $U^\ast$, given by Eq.~\eqref{eq:mRPA}, predict these observables as a function of $U_\infty$, $\alpha$ and $\omega_p$? The central idea behind Eq.~\eqref{eq:mRPA} is that impurity models which have the same mRPA moment are similar in terms of their physics. Thus, the applicability of mRPA can be tested by plotting the chosen observables versus the mRPA moment. When these curves at finite $\alpha$ collapse onto the $\alpha=0$ curve, the mRPA effective model is appropriate, for that particular observable and parameter regime.

An example of this test is shown in Fig.~\ref{fig:halffilled:colormap:beta10}(b). The colored symbols and lines show the results at $\alpha=3$, i.e., a horizontal slice of Fig.~\ref{fig:halffilled:colormap:beta10}(a), with the restriction $\alpha \leq U_\infty$ to avoid the bipolaron regime. For reference, the black line shows $\alpha=0$. Additional results for other temperatures are shown in Appendix~\ref{app:temperature}.
For the double occupancy, the match between the colored and black curves is almost perfect, showing that the mRPA value of $U^\ast$ correctly predicts the double occupancy. The double occupancy is special since it appears explicitly in Eq.~\eqref{eq:mRPA}. Furthermore, the double occupancy and instantaneous interaction strength are conjugate variables in the thermodynamic sense. Effective modelling tends to perform best when the target observable and the variational parameter are conjugate variables~\cite{vanLoon16b}.

For the kinetic energy and Green's function, the mRPA works very well in the strongly correlated metallic regime, which is most relevant for DMFT and similar approaches, whereas deviations are clearly visible in the weakly correlated regime (large mRPA moment). 
This observation is somewhat surprising, since RPA is by itself not accurate in the strongly correlated regime. We find that the mRPA benefits from error cancellations as described in Appendix~\ref{sec:supp:whyrpa}).

The mRPA performs better at large $\omega_p$ (green) and at small $\alpha$, since the dynamic effects are generally weaker in these regimes. Finally, the comparison in Appendix~\ref{app:temperature} shows that at constant $\alpha$, the mRPA performs more poorly at higher temperature. This is what one might generally expect for effective modelling in statistical physics, since at low temperatures the effective model only needs to describe a few low-lying excitations of the full model, while it needs to capture the entire excitation spectrum at high temperature. 

\begin{figure}
\includegraphics{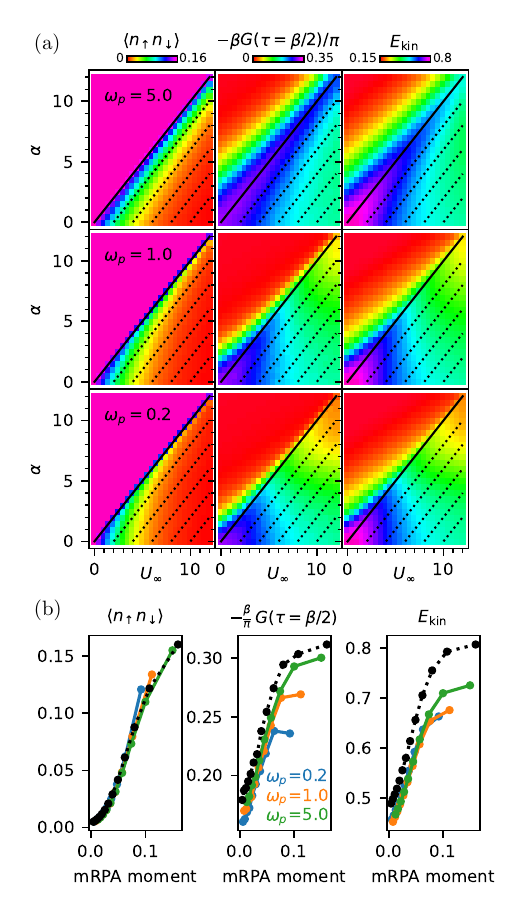}
\caption{Observables in the doped system at electron density $n\approx 0.8$, $\beta=10$. (a) color map (b) mRPA mapping at $\alpha=3$.
}
\label{fig:doped:colormap:beta10}
\end{figure} 
 
\section{Away from half-filling} 

Now, we move to the hole-doped system, $n\approx 0.8$. For the Hubbard model, this amount of doping is known to host the most challenging physics. Whereas the half-filled ($n=1$) case is dominated by the uniform metallic and Mott insulating phases, the doped regime shows a plethora of more complicated charge, spin and superconducting fluctuations~\cite{Xiao23}. The presence of holes changes the screening mechanisms and can therefore have a large impact on the determination of effective interactions~\cite{kapetanovic24}. 

Figure~\ref{fig:doped:colormap:beta10} shows that the double occupancy behaves qualitatively similarly to the half-filled case. The other observables show clear differences, with a dome-like structure appearing in the Green's function. Clearly, a single effective interaction cannot simultaneously describe these different $U_\infty$-$\alpha$ dependencies. 

This is further confirmed by looking at the observables versus the mRPA moments, Fig.~\ref{fig:doped:colormap:beta10}(b). The performance of mRPA for the double occupancy is still good, but deviations for the other observables are enhanced across the entire range of interactions. This is confirmed by the results at other temperatures and $\alpha$ shown in Appendix~\ref{app:doped}.

\begin{figure}
 \includegraphics{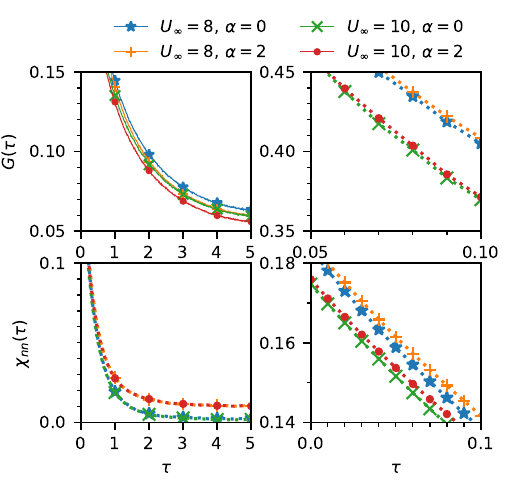}
 \caption{Green's function $G(\tau)$ and density-density correlation function $\chi(\tau)$ shown on long (left) and short (right) imaginary time scales, in the doped system ($n\approx 0.8$) for $\omega_p=0.2$ and $\beta=10$.}
 \label{fig:doped:dynamics}
\end{figure}

This breakdown of the effective $U^\ast$ modelling can be understood better by looking at the Green's function and density-density correlation function $\chi_{nn}(\tau)=\left\langle n(\tau) n(0) \right \rangle - \langle n \rangle^2$ in imaginary time, as shown in Fig.~\ref{fig:doped:dynamics}. For short times, both the Green's function and susceptibility are determined by $U_\infty$, while the susceptibility at long times, $\tau\approx \beta/2$, is mostly determined by $\alpha$. As also seen in Fig.~\ref{fig:doped:colormap:beta10}(a), the Green's function at $\tau=\beta/2$ actually decreases with $\alpha$ in this regime. In other words, dynamic screening increases the effective interaction strength for this observable in the doped system, in contrast to the half-filled system. None of these observables are governed by $U_\infty-\alpha=U(i\omega_0)$, which would show as a match between the blue and red line. A possible explanation for these features is the appearance of temporal phase separation to minimize the dynamic interaction energy, as discussed in Appendix~\ref{sec:supp:phaseseparation}.   

\section{Conclusion and discussion}

We have studied dynamic interactions in the single-impurity Anderson impurity model, and the ability of effective instantaneous interactions $U^\ast$ determined using mRPA~\cite{BoothMRPA} to capture the same physics. Overall, the mRPA performs very well and substantially better than the simple expression $U^\ast=U(i\omega_0)$, which is commonly used in the literature. Thus, our overall recommendation is to use the mRPA to estimate effective Hubbard interactions when needed. A benefit of the mRPA is that it can be performed at a negligible cost compared to the subsequent many-body calculation.

Still, although the effective modelling using mRPA works well overall, it is clear that there are also challenging regimes. We have found that effective models perform worse under doping, at higher temperature, in the weakly correlated regime and when the magnitude of the dynamic part is larger. The first two observations relate to the fact that effective modelling using a single parameter is harder when there are many low-lying excitations and the effective model needs to capture an entire ``energy landscape'', not just a few low-lying excitations. As such, these results show the limitations of effective modelling in general, not just of the specific mRPA recipe. In the weak correlation regime, our results speaks in support of $GW$+EDMFT~\cite{Tomczak12} or similar approaches that take into account the dynamic interactions in a consistent way.

In addition, the energy scale of the dynamic interaction, $\omega_p$, plays an important role. When $\omega_p$ is large, there is a decoupling of time scales and effective models are expected to work well~\cite{Casula12,aleryani2025}. Even the simple formula $U^\ast=U(i\omega_0)$ performs well in the limit of large $\omega_p$. This observation is familiar from the physics of phonon-driven superconductivity and the Schrieffer-Wolf transformation. On the other hand, the situation at small $\omega_p$ is more challenging, but even here the mRPA performs reasonably well and better than the simple formula $U^\ast=U(i\omega_0)$ (see Refs.~\cite{Cornaglia04,Sangiovanni05} for more discussion of the small $\omega_p$ limit). For cRPA calculations of $U(\omega)$ in solids, the bands that are integrated out are far away from the Fermi level and $\omega_p$ tends to be large compared to the bandwidth of the low-energy electronic structure (see Table~\ref{tab:values}), which supports the applicability of an effective model~\cite{Casula12}. Flat bands in twisted two-dimensional materials~\cite{bistritzer2011moire} are a potential exception to this. Although the low-energy bands themselves are very flat, the bands that are effectively integrated out also have very low energy and the background screening is therefore almost metallic~\cite{Pizarro19}.

Away from half-filling, there is a clear dichotomy between the double occupancy on one side and the kinetic energy and Green's function on the other side. The performance of the mRPA for the latter two observables is rather poor in the most challenging parameter regimes, which we believe could be caused by temporal phase separation, which cannot be represented in the effective model. The mRPA is defined to accurately describe the density matrix, which is not sensitive to these temporal fluctuations. The renormalization of the single-particle terms in the Hamiltonian~\cite{Casula12,Ayral17,Veld19} could help address this issue. In the context of the mRPA this would require introducing an additional constraint to determine the additional effective parameters.

The present benchmark involves most regimes of interest for single-orbital physics. An important open question is the role of dynamic screening in multi-orbital physics, especially in Hund's metals~\cite{georges2024hund}. In particular, it is plausible that the dynamic screening has a different impact on non-density-density terms in the Hamiltonian.

Finally, although the applicability of the cRPA in general is under active investigation~\cite{Honerkamp18,vanLoon21,chang2024downfolding,carta2025,Reddy25}, that is not the question being investigated here: treatment of effective interactions beyond cRPA will still lead to $U(\omega)$ and a need to reduce this to a single, instantaneous $U^\ast$. The current benchmark is purely about the mapping between $U(\omega)$ and $U^\ast$ and is agnostic to how $U(\omega)$ was obtained. At the same time, one obviously has to worry about propagation and possible cancellation of errors in multi-step, multi-approximation approaches. 
For this reason, simultaneous benchmarks of the entire downfolding procedure are important~\cite{chang2024downfolding,carta2025,kleiner2025quantummontecarloassessment}. Furthermore, it is worth pointing out that there are ways~\cite{changlani2015density,yu2020machine} of obtaining effective interactions $U^\ast$ as an effective parameter without ever constructing $U(\omega)$ from cRPA or similar approaches. However, first-principles approaches like cRPA have the advantage that they do not just provide numbers but also insight into the relevant physical processes responsible for the screening.

\acknowledgments

We acknowledge useful discussions with Tim Wehling and Mikhail Katsnelson. 
This work received support from Gyllenstiernska Krapperup's Foundation, from eSSENCE, a strategic research area for e-Science, grant number eSSENCE@LU 9:1 and the Swedish Research Council (Vetenskapsrådet, VR) under grant 2022-03090. The computations were enabled by resources provided by LUNARC, the Centre for Scientific and Technical Computing at Lund University through the projects LU 2025/2-47 and LU 2025/17-18.
M.R. acknowledges support from the Vidi ENW research programme of the Dutch Research Council (NWO) under the grant https://doi.org/10.61686/YDRHT18202 with file number VI.Vidi.233.077.

\appendix

\begin{figure}[t]
\includegraphics{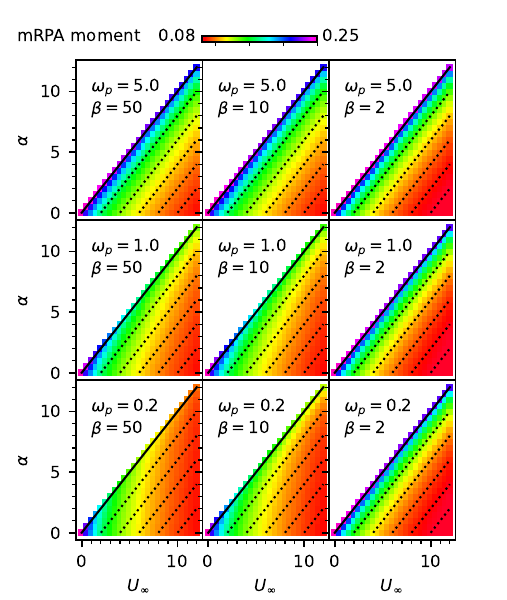}
 \caption{mRPA moments for several values of $\beta$ (columns) and $\omega_p$ (rows), at half-filling.  
 } 
 \label{fig:mrpa:moments:halffilled}
\end{figure}

\begin{figure*}[!t]
\includegraphics{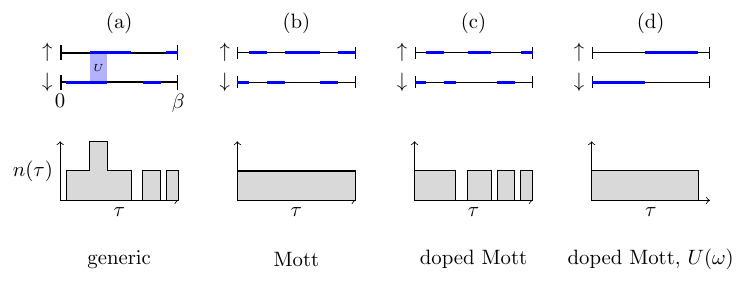} 
\caption{Segment picture of impurity models. (a) A generic time-evolution, the dark blue lines indicate when an orbital is filled and the shaded area denotes the Hubbard interaction when both orbitals are filled simultaneously. The average particle number per orbital is determined by the fraction of the line that is blue. (b) At half-filling and for large positive $U$, double occupancy is heavily suppressed. To reach half-filling while avoiding double occupancy, the impurity must also never be empty. (c) Upon hole-doping, empty time intervals appear. (d) With $U(\omega)$, filled segments attract and a temporal analog of phase separation appears.}
\label{fig:supp:timeCDW}
\end{figure*}

\section{mRPA moment}
\label{sec:app:mrpa}

The defining equation of the mRPA is Eq.~\eqref{eq:mRPA}. Here, we provide additional information on the precise definition (including normalization) and on practical details regarding the implementation. For the model in question, $\chi$ corresponds to the dynamic charge susceptibility of the impurity, i.e., $\chi(\tau)= \av{\hat{n}(\tau) \hat{n}(0)}-\av{\hat{n}}\av{\hat{n}}$ where $\hat{n}=\hat{n}_\up+\hat{n}_\dn$ is the total charge on the impurity. The value at $\tau=0$ thus corresponds to $\chi(\tau=0)=\av{\hat{n}\hat{n}}-\av{\hat{n}}\av{\hat{n}}=\av{\hat{n}^2_\up}+\av{\hat{n}^2_\dn}+2\av{\hat{n}_\up \hat{n}_\dn}-\av{\hat{n}}\av{\hat{n}}=\av{\hat{n}}-\av{\hat{n}}\av{\hat{n}}+2\av{\hat{n}_\up \hat{n}_\dn}$, where it was used that $\hat{n}_\sigma^2=\hat{n}_\sigma$ for fermions. Since we always consider the model at constant density, the only non-trivial part of the moment is the double occupancy $\av{\hat{n}_\up \hat{n}_\dn}$.

The double occupancy of the effective model, $\av{\hat{n}_\up \hat{n}_\dn}^\ast$, is directly related to its potential energy, $E_\text{pot}^\ast = U^\ast \av{\hat{n}_\up \hat{n}_\dn}^\ast$. Indeed, the central concept of the mRPA~\cite{BoothMRPA} is to reproduce the two-particle density matrix and thereby the energetics of the interaction. However, since $\av{\hat{n}_\up \hat{n}_\dn}^\ast$ would require the solution of the model with dynamic interactions, the mRPA instead uses the random phase approximation value of the double occupancy, i.e.,  $\chi(\tau=0)$. 

In the random phase approximation, one starts with the single-particle Green's function of the non-interacting impurity, $G^0(i\nu_n)=\left(i\nu_n-\epsilon_0+\mu-t^2 G^\text{bath}(i\nu_n)\right)^{-1}$. After Fourier transforming this to imaginary time, $G^0(\tau)$, the polarization or susceptibility of the non-interacting model is evaluated as $\Pi(\tau) = 2 G^0(\tau) G^0(\beta-\tau)$. Fourier transforming back to frequency gives $\Pi(i\omega_n)$. The RPA approximates the susceptibility of the interacting system as 
\begin{align}
 \chi^\text{RPA}(i\omega_n) = \frac{\Pi(i\omega_n)}{1+U(i\omega_n) \Pi(i\omega_n)}.
\end{align}
Finally, Fourier transforming back to imaginary time and evaluating at $\tau=0$ gives the mRPA moment
\begin{align}
 \text{mRPA moment} = \frac{1}{2} [\chi^\text{RPA}(\tau=0) - \av{n}(1-\av{n})], \label{eq:mrpa:moment}
\end{align}
where the factor $\frac{1}{2}$ and the subtraction of the constant part ensures that the moment is equal to the double occupancy.
At half-filling, the moment is 
$0.25$ in the non-interacting model, 0 for strong repulsive interactions and $0.5$ for strong attractive interactions. For $\av{n}=0.8$, the moment is $0.16$ for the non-interacting system, 0 for strong repulsive interactions and $0.4$ for strong attractive interactions.
 
\subsection{Analytical result for weak interactions}
The combination of Fourier transforms and inversions makes it difficult to analytically evaluate the mRPA moment, but in the regime of small $U(i\omega_n)$, we can Taylor expand $\chi^\text{RPA}(i\omega_n) \approx \Pi(i\omega_n)-\Pi(i\omega_n) U(i\omega_n) \Pi(i\omega_n)$. Thus, the mRPA condition becomes $\sum_n (U(i\omega_n)-U^{\ast,\text{mRPA}}) \, \Pi(i\omega_n)^2=0$. In other words, in this limit, $U^{\ast,\text{mRPA}}$ is the weighted average of $U(i\omega_n)$ with the weights provided by $\Pi(i\omega_n)^2$. In this view, using an instantaneous interaction corresponds to assuming that $U(i\omega_n)$ is constant in the frequency-range where the weights $\Pi(i\omega_n)^2$ are large. For larger interactions, higher orders in the Taylor expansion contribute and the relation between $U^{\ast,\text{mRPA}}$ and $U(i\omega_n)$ is non-linear.

\subsection{Numerical results for the mRPA mapping}

Figure~\ref{fig:mrpa:moments:halffilled} shows the mRPA moments as a function of temperature and $\omega_p$ for the half-filled case.
For any point in $(U_\infty,\alpha)$-parameter space, the mRPA effective interaction $U^\ast$ is obtained by following the isoline (constant color) down to $\alpha=0$. At large $\omega_p$, the isolines are straight and correspond to the simple recipe $U^\ast=U_\infty-\alpha$. At smaller $\omega_p$ and low temperature, for small $\alpha$ the isolines are still straight, but with a different slope. As discussed above,  in the small-$\alpha$ regime the effective interaction $U^\ast$ is a weighted average of $U(i\omega_n)$ and the changing slope shows that it is not just the low-frequency limit of $U(i\omega_n)$ that plays a role here. Staying at small $\omega_p$ and low temperature, at larger $\alpha$ the isolines are curved which shows that $U^\ast$ depends non-linearly on $\alpha$ according to mRPA. Based on the discussion above, this shows that higher-order terms in the Taylor expansion of the RPA are important. 

\section{Strongly correlated doped system in the segment picture}

\label{sec:supp:phaseseparation}

The impurity model studied here is solved numerically via Quantum Monte Carlo in the segment picture using the hybridization expansion~\cite{Gull:2011lr}. The combination of the hybridization expansion and the segment picture also provides a way to intuitively understand the basic physics of the impurity model with dynamic interactions. Figure~\ref{fig:supp:timeCDW} contains an illustration of the segment picture in different physical regimes.

We start with a model with instantaneous interaction $U$. In this work, we study an impurity with two spin-orbitals, $\up$ and $\dn$. For brevity, we call these spin-orbitals just orbitals in the following. The imaginary-time-evolution of the impurity model is described by segments (blue lines) indicating when each orbital is filled. When both orbitals are filled simultaneously, the impurity is doubly-occupied and there is an energy cost $U$, as indicated by the shaded area. Similarly, the chemical potential contribution to the grand potential is obtained from the total length of the segments. Thus, these two contributions to the grand potential and the corresponding observables have a clear visual interpretation in the segment picture. The kinetic energy, on the other hand, depends on the segments in a more complicated way. The Quantum Monte Carlo procedure involves performing a weighted average over different segment configurations.

Figure~\ref{fig:supp:timeCDW}(a) shows a generic imaginary time evolution, with some double occupancy and therefore a finite interaction energy. For large $U$, double occupancy is strongly suppressed, so that at half-filling and low temperature only configurations without any double occupancy, like Fig.~\ref{fig:supp:timeCDW}(b) survive. The Quantum Monte Carlo procedure therefore reduces to the averaging over half-filled, double-occupation-free configurations with different number and length of the segments, weighted by their kinetic energy (since the interaction energy is the same for all of them). The total density is time-independent and equal to 1.

By introducing holes into the system, there are necessarily some time intervals where both orbitals are empty, as illustrated in Fig.~\ref{fig:supp:timeCDW}(c), while it is still possible to avoid double occupancy entirely. Thus, the Quantum Monte Carlo procedure is similar to the undoped case in (b), but with more degrees of freedom, since there are both spin and charge fluctuations. 

When we introduce a dynamic charge interaction $U(\tau-\tau')$ instead of the instantaneous $U$, the regimes of (b) and (c) are affected differently. For the undoped case in (b), the charge density is constant and the same for all configurations. Thus, $\frac{1}{2} \int d\tau d\tau' U(\tau-\tau') n(\tau) n(\tau')$ only leads to a constant shift in the total energy. For the doped system, the situation is different. Fig.~\ref{fig:supp:timeCDW}(c) and (d) have the same double occupancy (zero, no overlapping segments) and particle number (total segment length), so they have the same interaction energy in an instantaneous interaction model, but they will have different energy in a dynamic interaction model (i.e., $U(\omega)$ or $U(\tau-\tau')$). 

In our model and in most solids, the dynamic part of the interaction is attractive, since it corresponds to the dynamic screening of the Coulomb interaction. The attractive interaction $U(\tau-\tau')$ decays roughly exponentially as a function of the imaginary time difference $\tau-\tau'$ and $1/\omega_p$ sets the time-scale of this decay. For small $\omega_p$, the attractive interaction decays slowly. It favors moving segments closer together, which in the most extreme case leads to a temporal phase separation between the $n(\tau)=1$ and $n(\tau=0)$ regions.  In Fig.~\ref{fig:doped:colormap:beta10}, this phenomenon is visible as the decrease of the number of segments $\av{k}=\beta E_\text{kin}$ with increasing $\alpha$ at small $\omega_p$ but not at large $\omega_p$.

\section{Temperature-dependence and additional data at half-filling}
\label{app:temperature}

\begin{figure}
 \includegraphics{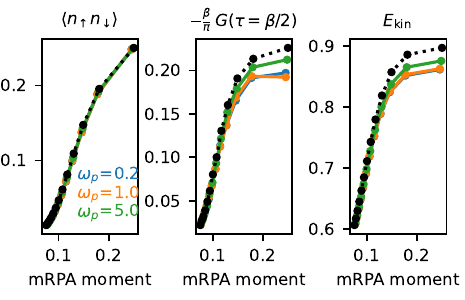}
 \includegraphics{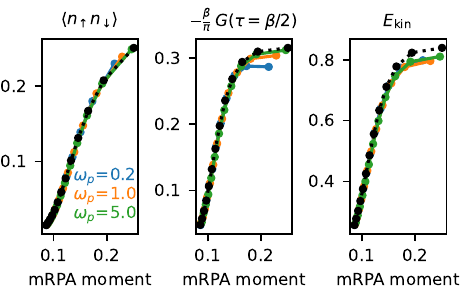}
 \includegraphics{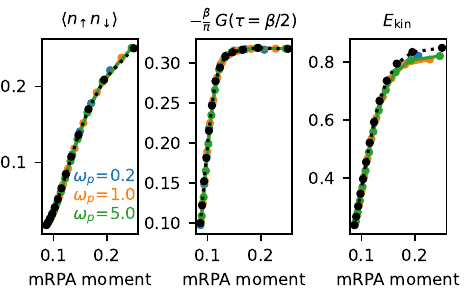} 
 \caption{How well does mRPA predict observables? Data for $\beta=2,10,50$ (top to bottom), $\alpha=1$, $\omega_p=0.2,1,5$ (colored lines). The black lines show the mRPA predictions. The mRPA works better for larger $\omega_p$, for larger $\beta$ and in the strongly correlated regime (small mRPA moment).}
 \label{fig:halffilled:mrpa:alpha1}
\end{figure}

\begin{figure}
 \includegraphics{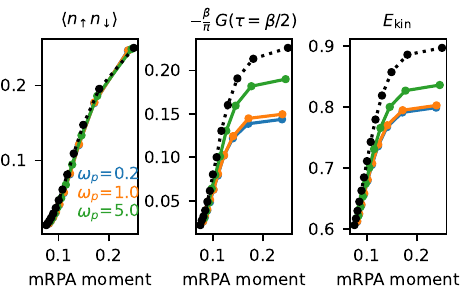}\\
 \includegraphics{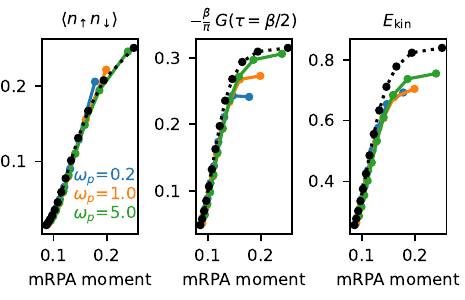}\\
 \includegraphics{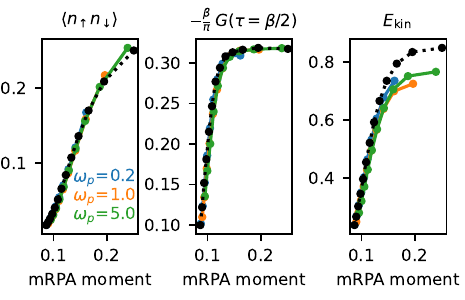} 
 \caption{Same as Fig.~\ref{fig:halffilled:mrpa:alpha1} but for the intermediate value $\alpha=3$.}
 \label{fig:halffilled:mrpa:alpha3}
\end{figure}

\begin{figure}
 \includegraphics{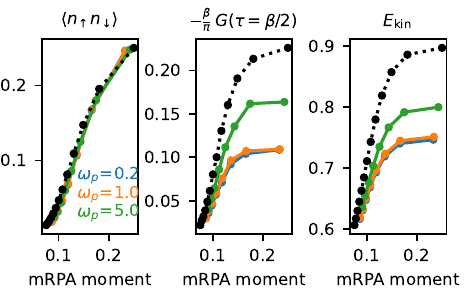}
 \includegraphics{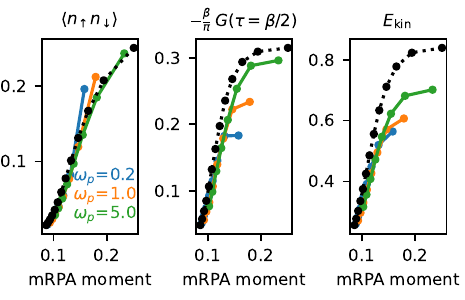}
 \includegraphics{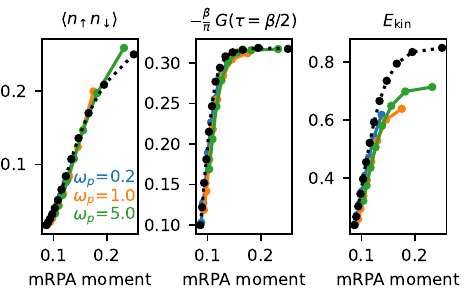} 
 \caption{Same as Fig.~\ref{fig:halffilled:mrpa:alpha1} but for the large value $\alpha=5$. Larger $\alpha$ is more challenging for the effective mapping.}
 \label{fig:halffilled:mrpa:alpha5}
\end{figure}

Figures~\ref{fig:halffilled:mrpa:alpha1}, \ref{fig:halffilled:mrpa:alpha3} and \ref{fig:halffilled:mrpa:alpha5} show the performance of mRPA at predicting observables for varying values of the dynamic interaction strength, $\alpha \in \{1,3,5\}$.
 
\section{Additional data for the doped system}
\label{app:doped}

\begin{figure}
\includegraphics{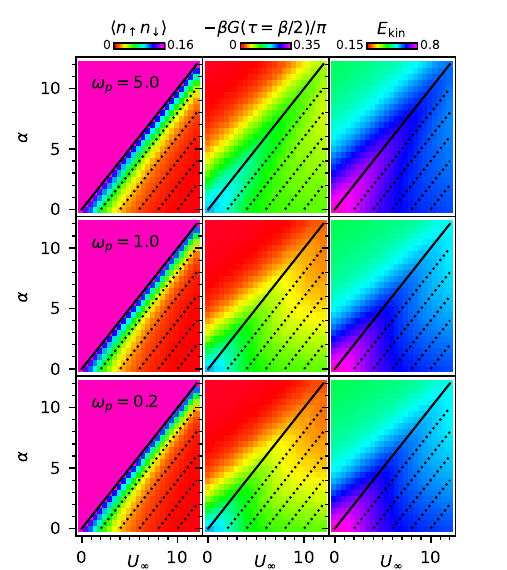}
\caption{Observables in the doped system at electron density $n\approx 0.8$, $\beta=2$. 
The black diagonal line shows $U_\infty-\alpha=U(i\omega_0)=0$ and the dotted lines are parallel to this line. The color scale for the double occupancy is restricted to show more detail in the region of interest.
}
\label{fig:doped:colormap:beta2}
\end{figure}

\begin{figure*}
 \includegraphics{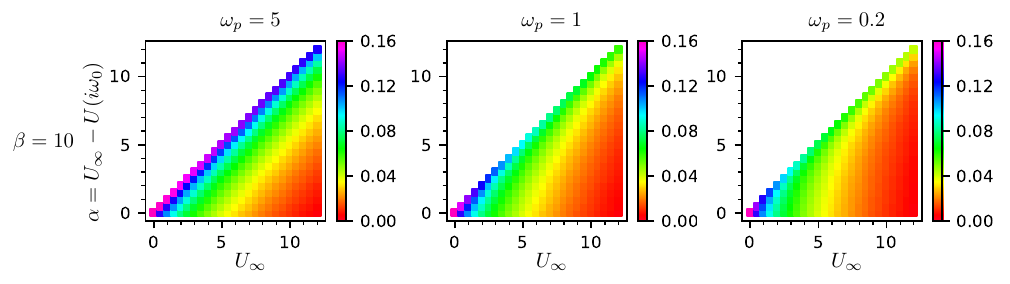}
 \caption{mRPA moments for several values of $\omega_p$ (columns), at $n\approx 0.8$. The doping-dependence is qualitatively similar to the half-filled case shown in  Fig.~\ref{fig:mrpa:moments:halffilled}.} 
 \label{fig:mrpa:moments:doped}
\end{figure*}

\begin{figure*}
 \includegraphics{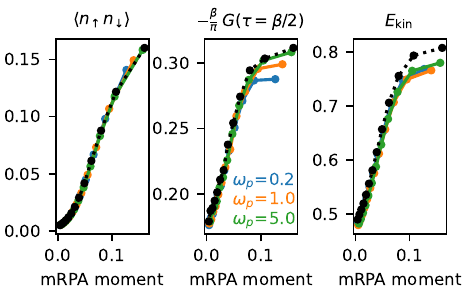}
 \hspace{0.6cm}
 \includegraphics{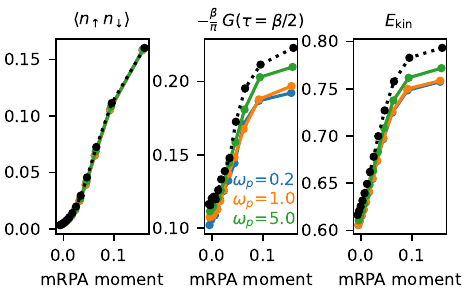}
 \includegraphics{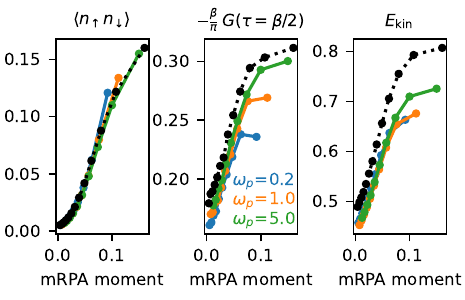}
 \hspace{0.6cm}
 \includegraphics{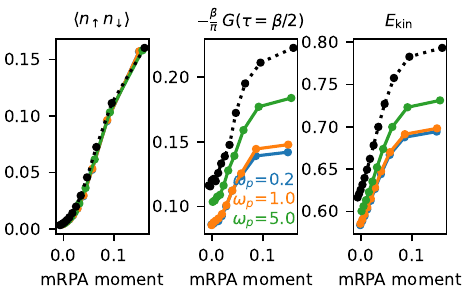}
 \includegraphics{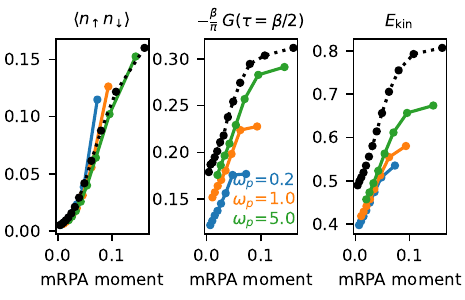} 
 \hspace{0.6cm}
 \includegraphics{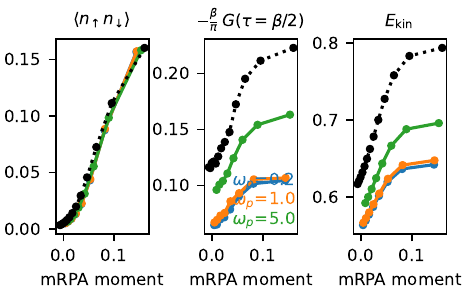} 
 \caption{How well does the mRPA predict observables in the doped system ($n\approx 0.8$)?. The left column shows $\beta=10$ while the right column shows $\beta=2$. The rows show $\alpha=1,3,5$ (top to    bottom). }
 \label{fig:doped:mrpa}
\end{figure*}

For the doped system, observables at $\beta=2$ are shown in Figure~\ref{fig:doped:colormap:beta2}, the mRPA moments at $\beta=10$ in Fig.~\ref{fig:mrpa:moments:doped} and the performance of the mRPA for predicting observables in Figure~\ref{fig:doped:mrpa}.

%\clearpage 
\begin{figure*}[t]
 \includegraphics{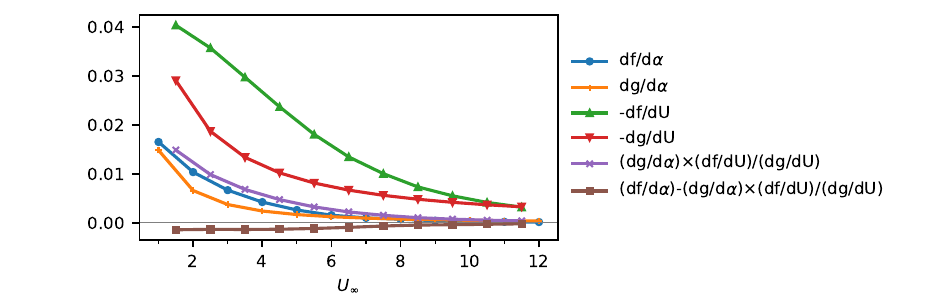}
 \caption{Estimate of the derivatives in Eq.~\eqref{eq:supp:error} using finite differences, at half-filling and $\omega_p=0.2$, $\beta=10$. Note that small $\omega_p$ is most challenging for effective mappings, as discussed in the main text. Here, $f$ is the exact double occupancy calculated using CTSEG while $g$ is the RPA estimate of the double occupancy.}  
 \label{fig:supp:exactvsrpa}
\end{figure*}

\section{Why does mRPA work? Small-$\alpha$ perturbation theory}
\label{sec:supp:whyrpa}

The colormaps in this manuscript illustrate the determination of an effective model in the $(U_\infty,\alpha)$ plane, keeping all other parameters (temperature, $\omega_p$, filling) constant. To simplify the notation, we write $f(U_\infty,\alpha)$ for the exact double occupancy as a function of $U_\infty$ and $\alpha$ and $g(U_\infty,\alpha)$ for the RPA estimate of the double occupancy. The goal of effective modeling is to find a mapping $(U_\infty,\alpha)\mapsto (U^\ast,0)$ such that $f(U_\infty,\alpha)\approx f(U^\ast,0)$. We can perform a Taylor expansion in $\alpha$ and $U_\infty-U^\ast$, with the expectation that $U_\infty-U^\ast$ will turn out to be linear in $\alpha$ at lowest order (confirmed explicitly for mRPA below),
\begin{align}
 f(U_\infty,\alpha)-f(U^\ast,0) = (U_\infty-U^\ast) \frac{df}{dU} + \alpha \frac{df}{d\alpha} + O(\alpha^2). \label{eq:supp:taylorf}
\end{align}
Note that we are not assuming small $U$ here, so this equation is valid even in the correlated regime. A similar Taylor expansion holds for $g$,
\begin{align}
 g(U_\infty,\alpha)-g(U^\ast,0) = (U_\infty-U^\ast) \frac{dg}{dU} + \alpha \frac{dg}{d\alpha} + O(\alpha^2). \label{eq:supp:taylorg}
\end{align}
In the mRPA, the condition $g(U_\infty,\alpha)-g(U^\ast,0)=0$ defines $U^\ast$, so in the linear regime we indeed have a linear dependence of $U_\infty-U_\ast$ on $\alpha$,
\begin{align}
 (U_\infty-U^\ast) \frac{dg}{dU} + \alpha \frac{dg}{d\alpha}\overset{\text{mRPA}}{=}0.
\end{align}
Putting this result into Eq.~\eqref{eq:supp:taylorf} gives
\begin{align}
  f(U_\infty,\alpha)-f(U^\ast,0) = \alpha \left[\frac{df}{d\alpha}- \frac{ \left(\frac{df}{dU}\right)\left(\frac{dg}{d\alpha}\right) }{ \left(\frac{dg}{dU}\right) } \right]. \label{eq:supp:error}
\end{align}
Thus, the performance of the mRPA for predicting the double occupancy, in the small-$\alpha$ and arbitrary $U$ regime is determined by the combination of derivatives in square brackets. From our calculations, it is possible to estimate these derivatives using finite differences. The result is shown in Fig.~\ref{fig:supp:exactvsrpa}. There are clear differences between $df/d\alpha$ and $dg/d\alpha$ (blue and orange), and between $df/dU$ and $dg/dU$ (green and red), because the RPA is not a good estimate of the true double occupancy in the strongly correlated regime. However, looking at the relevant combination of Eq.~\eqref{eq:supp:error} (brown line) shows that the total error is quite small, especially in the strongly correlated regime. The reason for this is an error cancellation between the RPA estimates of the derivative with respect to $U$ and $\alpha$ (both underestimated) as well as an overall decrease in the dependence of the double occupancy on $\alpha$. 

Finally, note that according to Eq.~\eqref{eq:supp:taylorf}, the simple formula $U^\ast=U_\infty-\alpha=U(i\omega_0)$ would work well when $df/dU=-df/d\alpha$ (blue line is equal to green line), which is clearly not the case here.

Thus, we conclude that the mRPA is not exact for the prediction of the double occupancy in the small-$\alpha$ limit, but it is rather close due to a cancellation of errors which is not present in the simple $U^\ast=U(i\omega_0)$ formula. 

\bibliography{references}

\end{document}